\documentclass[preprint,11pt,square,numbers,sort&compress,comma]{elsarticle}
\usepackage{footnote}
\usepackage{booktabs}
\usepackage{multirow}
\usepackage{algorithm}
\usepackage{algorithmic}
\usepackage{threeparttable}
\usepackage{amssymb}
\usepackage{amsmath}
\usepackage{multirow}
\usepackage{graphicx}
\usepackage{pdflscape}
\usepackage{geometry}
\usepackage{setspace}
\usepackage{longtable}
\usepackage{dcolumn}
\usepackage{bm}
\usepackage{color}
\usepackage{soul}

\journal{Advances in Engineering Software}

\begin{document}

\begin{frontmatter}

\title{Rapid online solution of inverse heat transfer problem by ANN-based extended Kalman smoothing algorithm}


\author[mymainaddress]{Xinxin Zhang}
\author[mymainaddress]{Dike Li}
\author[mymainaddress]{Jianqin Zhu}
\author[mymainaddress]{Zhi Tao}
\author[mymainaddress]{Lu Qiu \corref{mycorrespondingauthor}}
\cortext[mycorrespondingauthor]{Corresponding author}
\ead{qiulu@buaa.edu.cn}

\address[mymainaddress]{National Key Laboratory of Science and Technology on Aero-Engine Aero-Thermodynamics,\\School of Energy and Power Engineering, \\Beihang University, \\Beijing 100191, PR China}

\doublespacing

\begin{abstract}
Digital twin is a modern technology for many advanced applications. To construct a digital twin of a thermal system, it is required to make online estimations of unknown time-varying boundary conditions from sensor measured data, which needs to solve inverse heat transfer problems (IHTPs). However, a fast and accurate solution is challenging since the measured data is normally contaminated with noise and the traditional method to solve IHTP involves significant amount of calculations. Therefore, in this work, a rapid yet robust inversion algorithm called ANN-based extended Kalman smoothing algorithm is developed to realize the online prediction of desired parameter based on the measurements. The fast prediction is realized by replacing the conventional CFD-based state transfer models in extended Kalman smoothing algorithm with pre-trained ANN. Then, a two-dimensional internal convective heat transfer problem was employed as the case study to test the algorithm. The results have proved that the proposed algorithm is a computational-light and robust approach for solving IHTPs. The proposed algorithm can achieve estimation of unknown boundary conditions with a dimensionless average error of 0.0580 under noisy temperature measurement with a standard deviation of 10 K with a drastic reduction of computational cost compared to the conventional approach. Moreover, the effects of training data, location of sensor, future time step selection on the performance of prediction are investigated.
\end{abstract}


\end{frontmatter}

\section*{Nomenclature}
$%
\begin{array}{ll}
k_{c} & \text{Coefficient of thermal conductivity (W/(m$\cdot$K))} \\
\rho & \text{density (kg/m$^3$)} \\
C_{p} & \text{isobaric specific heat capacity (J/(kg$\cdot$K))} \\
L & \text{length (m)} \\
h & \text{height (m)} \\
q & \text{wall heat flux (W/m$^{2}$)} \\
X & \text{the horizontal distance in the X-axis (m)}\\
Y & \text{the vertical distance in the Y-axis (m)}\\
T & \text{temperature (K)} \\
u & \text{flow velocity (m/s)} \\
m & \text{sensor noise level (K) } \\
\Gamma  & \text{state transfer operator} \\
H & \text{measurement operator} \\
\omega & \text{state transfer model noise vector} \\
\nu  & \text{measurement model noise vector}\\
Q& \text{the covariance matrix of state transfer model noise} \\
R& \text{the covariance matrix of measurement noise} \\
x & \text{the state vector} \\
y & \text{the measurement vector} \\
\hat{x}&\text{the predicted state vector } \\
\hat{y}&\text{the predicted measurement vector} \\
\tilde{x}&\text{the corrected state vector } \\
\tilde{y}&\text{the corrected measurement vector} \\
\hat{P} & \text{the predicted covariance matrix of the state vector} \\
\tilde{P} & \text{the corrected covariance matrix of the state vector} \\
F  &  \text{Jacobi matrix}\\
\varepsilon & \text{infinitesimal} \\
e & \text{unit vector} \\
K & \text{Kalman gain} \\
\tilde{x}^{\prime}&\text{the smoothed state vector } \\
\tilde{P}_{k}^{\prime}&\text{the smoothed covariance matrix } \\
n_{f}&\text{future time step} \\
G&\text{Kalman gain in smoothing algorithm } \\
AE&\text{dimensionless average error } \\
\\\text{Subscripts} & \text{} \\
in & \text{inlet parameter} \\
k & \text{the current time step} \\
i,j & \text{mesh node index} \\
s & \text{sensor parameter} \\
\end{array}
$

\doublespacing
\section{Introduction}
In many transient convective heat transfer problems, the unknown time-varying thermal boundary conditions (BCs) are difficult to be measured directly\cite{Edwards1993}, yet whose online estimation is essential for improving the stability and performance of the thermal system in various engineering applications, such as aerospace thermal protection \cite{2}, chip cooling\cite{3}, metallurgical reactors\cite{4,5} and food engineering\cite{6}. The inverse heat transfer problems (IHTPS) have been developed to estimate the unknown time-varying boundary conditions from the interior temperature fields. With the aid of the IHTP algorithm, it is possible to reconstruct  the thermal boundary condition as well as the full temperature field from the online data measured by temperature sensors installed somewhere in the fluid domain. The technology is also known as digital twin.

However, the traditional methods to solve IHTPs faces the following challenges and drawbacks. Firstly, it is difficult to obtain relatively stable and accurate solutions under noisy input data due to the inherent ill-posedness of IHTPs\cite{7,Patel2021}. Secondly, it is difficult to invert the unknown heat flux from delayed sensor measurements since the thermal disturbance decays at the downstream due to the diffusive nature of heat transfer\cite{8}. Thirdly, the traditional algorithm to solve IHTPs is a computation-intensive and time-consuming process\cite{9}, which is difficult to be applied to online estimations.

The solutions for estimating time-varying BCs can be divided into two categories, namely, the whole domain algorithm and the sequential algorithm. The whole domain algorithm, by definition, is an offline method to calculate the unknown boundary conditions when all the time measurements in full time period are available. The algorithm transforms the IHTPs to an optimization problem, iteratively calculating the forward heat transfer process to minimize the error between the assumed and exact value of the unknown BCs by traditional optimization algorithms like Levenberg-Marquardt Method\cite{10} or heuristic algorithms like repulsive particle swarm algorithm\cite{11}, ant colony optimization, and cuckoo search algorithm\cite{12}.

The sequential algorithms, on the other hand, work in the online mode. With continuously measured instantaneous temperature, the real-time or near real-time estimation of the unknown BCs can be achieved. Many sequential inversion methods have been brought up, including the sequential function specification method (SFSM)\cite{13}, dynamic matrix control method\cite{14}, multiple model adaptive inverse (MMAI) method\cite{15}, artificial neural network (ANN) algorithm\cite{9,Tamaddon2020,Florent2021, 16, 17}, and digital filter (DF) approach\cite{2, 4, 8, 18,19,20,21,22,23}, etc.

Beck J.V\cite{13} firstly introduced the concept of future measurement into IHTPs, which tackles the sensing delay issues to ensure a stabilized solution without time-lag or distortion. In order to reduce the interference caused by the noises, the digital filtering approach\cite{2, 4, 8, 18,19,20,21,22,23} are employed. Owing to its statistical trade-off between the sensor measurement and model prediction towards a smoothed result, the Kalman filtering approach is relatively more robust under noisy input data, which serves as a valuable tool for tackling the ill-posed problem\cite{7}. Scarpa.F and Milan.G\cite{24} employed the Kalman filtering (KF) technique to solve a linear one-dimensional heat conduction problem, which shows the anti-interference ability of KF algorithms. The algorithm could be coupled with Rauch-Tung-Strieber (RTS) smoothing, which utilizes future time measurement to reduce the time lag in the prediction results. Although standard KF technique can be used to tackle liner IHTPs efficiently, it does not show pleasant for nonlinear IHTPs. Therefore, the extended version of Kalman filtering (EKF) technique\cite{25}, unscented Kalman smoothing algorithm\cite{26}, and other KF related methods\cite{4, 8, 18, 21}  are developed  to deal with nonlinear IHTPs. Although the KF-related algorithms work well under noisy environment, but all of them share a common drawback, which is the prohibitively high computational cost caused by repetitive CFD forward calculations required in the sensitivity analysis.

Other algorithm, such as the artificial neural network (ANN) algorithm, with its powerful mapping ability\cite{28} and high computational efficiency, may help to reduce computational cost. For instance, Najafi.H et al\cite{16,17} utilized ANN models to directly correlate unknown BCs with sensor-measured temperatures, while Huang.S et al\cite{9} utilized well-trained ANN structures as the rapid forward solver, coupled with inverse algorithm, indirectly realizing the online estimation of unknown BCs. Both works demonstrate that the ANN algorithm could accelerate the computation of IHTPs with satisfactory accuracy. However, the major drawbacks of ANN algorithm are that, the training of ANN requires vast amount of dataset under high representation load, and it tends to over-fitting when the sensor measurements are contaminated with noises.

To summarize, the KF-related approach is a robust method when measurement noise is non-negligible, but its high computational cost limits its applications towards online estimation task. The ANN models, on the other hand, as the universal approximators, could be employed to improve the computational efficiency of KF-related algorithm. Thus, in this work, we try to combine the extended Kalman smoothing algorithm and ANN algorithm to establish a rapid yet robust solution to IHTPs. Moreover, to reduce the redundant calculation of sensitivity analysis requires in the traditional EKF approach, we developed a reduced form of EKF state vector with the aid of ANN model to further improve the computational efficiency. A two-dimensional convective heat transfer problem is selected as the case study for the implementation and evaluation of the proposed algorithm.

\section{Inversion procedure by extended Kalman smoothing}
\subsection{The benchmark problem}
\begin{figure}[h]
\centering
\includegraphics[scale=1]{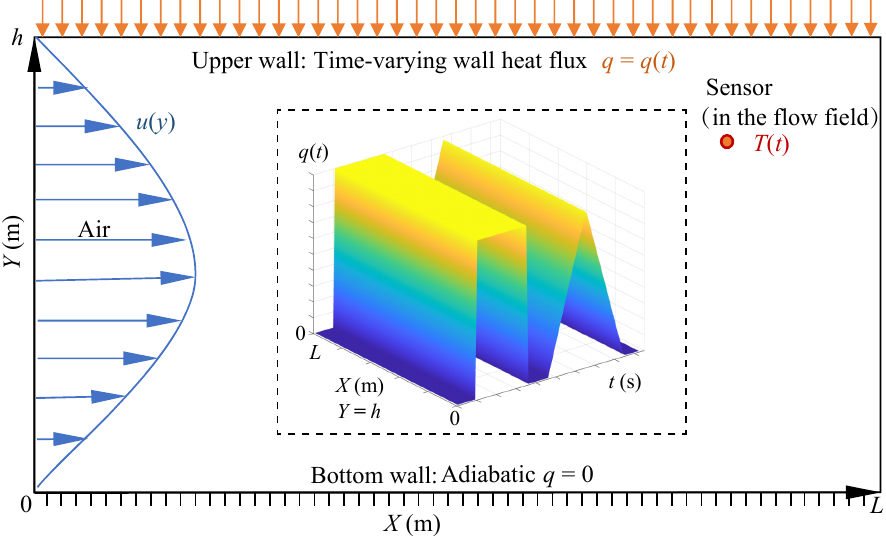}
\caption{The schematic of the benchmark two-dimensional heat convection inverse problem.The Inset shows an example of time-varying heat flux applied on the upper boundary. } \label{fig1}
\end{figure}

The case study is showed in Fig. \ref{fig1}, which describes a convective heat transfer problem on a two-dimensional rectangular region. A sensor is placed in the field, providing us with real-time temperature measurement while a time-varying heat flux, as the unknown BCs, is applied on the upper boundary. The main objective of our proposed algorithm is to estimate this unknown $q(t)$ in online mode by utilizing the continuously measured sensor temperatures $T(t)$.

The bottom boundary is considered to be adiabatic. On the left boundary, fully developed air with an initial temperature $T_{i n}=300 \mathrm{~K}$ and an average inlet velocity $u_{m}=0.033 \mathrm{~m} / \mathrm{s}$ passes through the region.
The sensor location is set to be $(0.820,0.089)$ and other detailed parameters are showed in Table.\ref{tab1}. Then, we give the governing equation for the above problem,
\begin{equation}\label{eq.1}
\rho C_{p} \frac{\partial T(X, Y, t)}{\partial t}+\rho C_{p} u(Y) \frac{\partial T(X, Y, t)}{\partial X}=k_{c} \frac{\partial^{2} T(X, Y, t)}{\partial Y^{2}}
\end{equation}
where the boundary conditions and initial conditions are.
\begin{equation}\label{eq.2}
\begin{gathered}
k_{c} \frac{\partial T(X, Y=h, t)}{\partial Y}=q(t) \\
k_{c} \frac{\partial T(X, Y=0, t)}{\partial Y}=0 \\
u(X=0, Y, t)=u(Y)=6 u\left[\frac{Y}{h}\left(1-\frac{Y}{h}\right)\right] \\
T(0, Y, t)=T_{\text {in }}
\end{gathered}
\end{equation}
\begin{table}[htbp]
 \caption{\label{tab1}Detailed parameters for the numerical example.}
 \centering
\begin{tabular}{ccc}
\hline  Symbol & Quantity & Value \\
\hline$k_{c}$ & Coefficient of thermal conductivity & $0.243 \mathrm{~W} /(\mathrm{m} \cdot \mathrm{K})$ \\
$\rho$ & Air density & $1.29 \mathrm{~kg} / \mathrm{m}^{3}$ \\
$C_{p}$ & Specific heat capacity & $1005 \mathrm{~J} /(\mathrm{kg} \cdot \mathrm{K})$ \\
$L$ & Length & $1 \mathrm{~m}$ \\
$h$ & Height & $0.1 \mathrm{~m}$ \\
\hline
\end{tabular}
\end{table}

\subsection{Inversion procedure by extended Kalman smoothing }
\begin{figure}[h]
 \raggedleft
\includegraphics[scale=0.92]{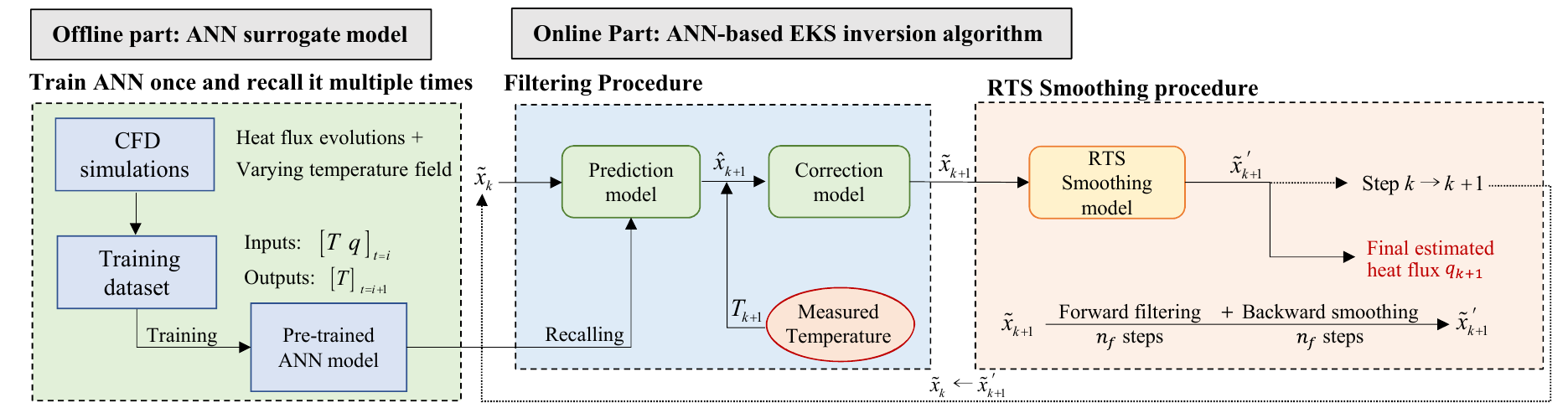}
\caption{The procedure of estimating time-varying heat flux at $t = k+1$ by the proposed algorithm} \label{fig2}
\end{figure}

In this work, the IHTPs is solved by the ANN-based extended Kalman smoothing algorithm (ANN-EKS) which composes of two main parts, ANN-based forward solver and extended Kalman smoothing algorithm. The entire procedure is summarized as Fig. \ref{fig2}. The extended Kalman smoothing algorithm gives estimation of the unknown boundary conditions step by step and is accelerated based on the fast prediction of local temperature field and sensitivity analysis made by ANN-based forward solver.

The extended Kalman smoothing algorithm can be further separated into two sections, namely, the extended Kalman filtering (EKF) algorithm and the Rauch-Tung-Strieber (RTS) smoothing algorithm. The RTS smoothing algorithm enables the EKF algorithm to include future measurement into the estimation of unknown heat flux, which can address the sensing delay issue of IHTPs.

\subsubsection{Extended Kalman filtering}
As aforementioned, the IHTPs is ill-posed by nature\cite{7}, which means small disturbance of the input data may cause large error in the output result. To solve this challenge, the extended Kalman filtering algorithm is employed, which uses the state-space representation to describe this two-dimensional convective heat transfer process and statistically quantifies the model error and the measuring noises. The algorithm compromises between the original prediction and the noisy measurements so that a more precise estimation could be made.

At first, the benchmark problem needs to be described under the state-space representation, whose nonlinear forms are showed in  Eqs. \ref{eq.3},
\begin{equation} \label{eq.3}
\begin{aligned}
&x_{k+1}=\Gamma\left(x_{k}, z_{k}\right)+\omega_{k+1} \\
&y_{k+1}=\mathrm{H}\left(x_{k+1}\right)+v_{k+1}
\end{aligned}
\end{equation}
where $x$ is the state vector of the system describing its current state, $y$ is the measurement vector, and $z$  is the input of the system. The nonlinear operator $\Gamma$ transfers the system state from current time step to the next and the measurement operator H maps the state vector to the measurement vector. The $\omega$ and $\nu$ are respectively the state transfer model error and measurement noise, which are assumed to be independent zero-mean Gaussian noises with covariance matrix $Q$ and $R$.

In order to apply the EKF algorithm for solving IHTPs, proper state variables need to be selected to fully describe the state of given thermal system. In literature\cite{24,25,26}, state vector  $x_{k}$ in the following form was employed,

\begin{equation}\label{eq.4}
x_{k}=\left[\begin{array}{llll}
T_{k}^{1} & T_{k}^{2} \ldots T_{k}^{i} \ldots T_{k}^{N} & q_{k}
\end{array}\right]^{T}
\end{equation}
where $T_{k}^i(i=1,2,...,N)$ is the temperature of node $i$ at time $k$, and $N$ is the amount of mesh nodes used in the numerical computation of the benchmark problem, representing the temperature field in discretized form. It is noteworthy that, instead of representing the input $z$, the unknown boundary heat flux $q_k$ was augmented into the state vectors in order to  estimate the unknown heat flux.

In IHTPs, the measurement vector $y$ in Eqs. \ref{eq.3} represents the sensor-measured temperature and the nonlinear measurement H operator is then reduced to a linear matrix $H$ showed as follows,
\begin{equation}\label{eq.5}
H=\left[\begin{array}{lllll}
0 & 0 & \cdots & \underbrace{1}_{i=n_{s}} \cdots & 0
\end{array}\right]_{1, N+1}
\end{equation}
where the node index $n_s$ corresponds to the sensor location in the temperature field.

The EKF approach estimates the next-time-step unknown boundary heat flux in two phases, the prediction phase and the correction phase.

In the prediction phase, the EKF approach predict the unknown state vector as well as its probability distribution for the next-time step, which are realized by updating the means and covariance matrix of state vector $x_{k+1}$ based on the current $x_{k}$ and the state space model of Eqs. \ref{eq.3}.

To calculate the means and covariance matrix of $x_{k+1}$, the EKF approach further linearize this problem by approximating $\Gamma\left(x_{k}\right)$ with first-order Taylor expansion at $\tilde{x}_{k}$,
\begin{equation}\label{eq.7}
\Gamma\left(x_{k}\right)=\Gamma\left(\tilde{x}_{k}\right)+\left.\frac{\partial \Gamma}{\partial x_{k}}\right|_{x_{k}=\tilde{x}_{k}}\left(x_{k}-\tilde{x}_{k}\right)+o\left(x_{k}-\tilde{x}_{k}\right)
\end{equation}
where $\tilde{x}_{k}$ is the corrected estimation result at last time step $k$, and $\frac{\partial \Gamma}{\partial x_{k}}$ is the Jacobi matrix of the state vector, denoted as $F_{k}$.
\begin{equation}\label{eq.8}
F_{k}=\left.\frac{\partial \Gamma}{\partial x_{k}}\right|_{x_{k}=\tilde{x}_{k}}=\left[\begin{array}{ccc}
\frac{\partial \Gamma_{1}}{\partial x_{k, 1}} & \cdots & \frac{\partial \Gamma_{1}}{\partial x_{k, N+1}} \\
\vdots & \vdots & \vdots \\
\frac{\partial \Gamma_{N+1}}{\partial x_{k, 1}} & \cdots & \frac{\partial \Gamma_{N+1}}{\partial x_{k, N+1}}
\end{array}\right]_{(N+1) \times(N+1)}
\end{equation}
It can be numerically calculated by the following means,
\begin{equation}\label{eq.9}
\frac{\partial \Gamma_{i}}{\partial x_{k, j}} \approx \frac{\Gamma_{i}\left(x_{k}+e_{j} \varepsilon x_{k, j}\right)-\Gamma_{i}\left(x_{k}-e_{j} \varepsilon x_{k, j}\right)}{2 \varepsilon x_{k, j}}
\end{equation}
where $e_{j}=[0 \cdots {1}_{j} \cdots 0]_{N+1}^{T}$ and $\varepsilon$ is an  infinitely small quantity  set to be $10^{-4}$ .

Since the algorithm is a recursive procedure, the   corrected state vector $\hat{x}_{k}$ and its covariance matrix $\tilde{P_{k}}$ at the time step $k$ have already been obtained, which will be utilized to calculate the next-time-step predicted state vector $\hat{x}_{k+1}$ (means of ${x}_{k+1}$) and its covariance matrix $\hat{P}_{k+1}$  as follows,
\begin{equation}\label{eq.10}
\begin{gathered}
\hat{x}_{k+1}=\Gamma\left(\tilde{x}_{k}\right) \\
\hat{y}_{k+1}=H \hat{x}_{k+1} \\
\hat{P}_{k+1}=F_{k} \tilde{P_{k}} F_{k}^{T}+Q
\end{gathered}
\end{equation}
where $\hat{y}_{k+1}$ is the predicted measurement temperature at time step $k+1$ .

In the correction phase, the Kalman gain is calculated as follows, which can be considered as the confidence level ratio between the model prediction and sensor measurements.
\begin{equation}\label{eq.11}
K=\hat{P}_{k+1} H^{T}\left(H \hat{P}_{k+1} H^{T}+R\right)^{-1}
\end{equation}

In the inversion procedure, the sensor-measured temperature $y_{k+1}$ , despite being contaminated with noises, provides true physical information. The difference bewteen sensor-measured $y_{k+1}$ and the model prediction $\hat{y}_{k+1}$ is multiplied by the Kalman gain to form the correction value $\Delta \tilde{x}$.

\begin{equation}\label{eq.12}
\begin{gathered}
\Delta \tilde{x}=K\left(y_{k+1}-\hat{y}_{k+1}\right)\\
\tilde{x}_{k+1}=\hat{x}_{k}+\Delta \tilde{x} \\
\tilde{P}_{k+1}=\hat{P}_{k+1}-K H \hat{P}_{k+1}
\end{gathered}
\end{equation}
where the corrected state vector $\tilde{x}_{k+1}$ and its covariance matrix $\tilde{P}_{k+1}$ will be also used to make predition for time step $k+2$ as showed in Eqs. \ref{eq.10}, forming a complete recursive procedure.
\subsubsection{RTS smoothing }

The EKF approach is a real-time inversion algorithm, which utilizes the current available data to estimate the unknown boundary heat flux. However, due to the diffusive nature of heat transfer process, the thermal response at sensor location lags behind the changing boundary heat flux, which is difficult to be captured by EKF approach itself.

In this work, a fixed interval smoothing algorithm called the Rauch-Tung-Strieber (RTS) smoothing technique is employed to include the data of future time measurement into the algorithm for better estimations.
\begin{figure}[h]
\centering
\includegraphics[scale=1]{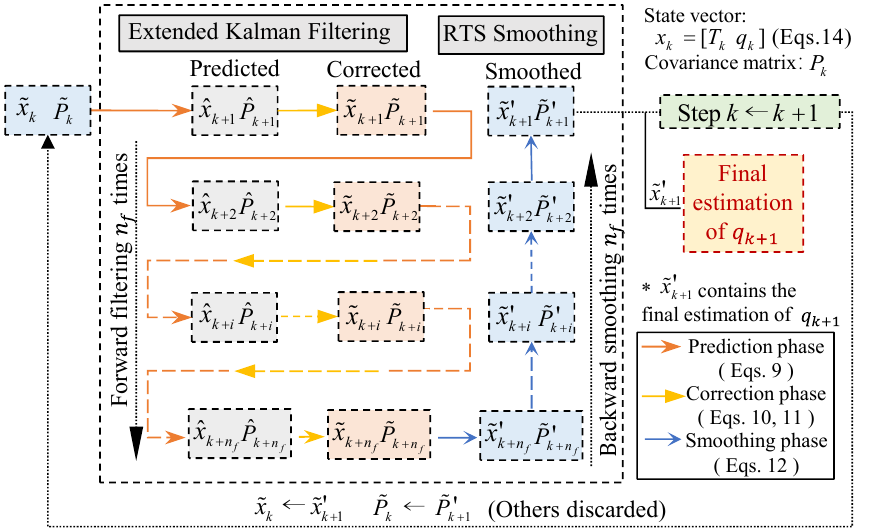}
\caption{The procedure of RTS smoothing at time step $k+1$} \label{fig3}
\end{figure}
As showed in Fig. 3, the RTS algorithm loops Kalman filtering procedures forwardly by $n_{f}$ times and then slides back by $n_{f}$ times to obtain a smoothed result. In the forward filtering procedures, we have already calculated the predicted $\hat{x}_{k}$, the corrected estimation $\tilde{x}_{k}$, the corresponding predicted error covariance $\hat{P}_{k}$, the corrected $\tilde{P}_{k}$, and the Jacobi matrix $F_{k}$ at time step $k\left(k=k_{0} \ldots k_{0}+n_{f}\right)$. Thus, the backward recursion procedures can be proceeded as follows,

\begin{equation}\label{eq.13}
\begin{gathered}
G_{k}=\tilde{P}_{k} F_{k}^{T}\left(\hat{P}_{k+1}\right)^{-1} \\
\Delta{x}^{\prime}=G_{k}\left(\tilde{x}_{k+1}^{\prime}-\tilde{x}_{k+1}\right)\\
\tilde{x}_{k}^{\prime}=\tilde{x}_{k}+\Delta{x}^{\prime}\\
\tilde{P}_{k}^{\prime}=\tilde{P}_{k}+G_{k}\left(\tilde{P}_{k+1}^{\prime}-\hat{P}_{k+1}\right) G_{k}^{T}
\end{gathered}
\end{equation}
where $\tilde{x}_{k}^{'}$,  $\tilde{x}_{k+1}^{'}$  and $\tilde{P}_{k}^{\prime}$ are the smoothed results and $G_{k}$ represents the RTS version of Kalman gain.

\section{The ANN-based rapid state transfer model }
In traditional KF-related approaches, the state transfer is usually achieved by CFD simulations, whose computation may take longer than the physical time in complex flow and heat transfer problems, thus being unsuitable for online inversion.

Alternatively, the artificial neural network is employed as a surrogate model for temperature field prediction by CFD, which can significantly reduce computational cost while holds certain level of accuracy\cite{28}.

More importantly, given that the ANN prediction does not require the information of entire temperature field, it allows to reduce the dimension of state vector in the EKS algorithm, which can tremendously reduce the computational cost and eventually realize online predictions.

\subsection{The general form of ANN-based state transfer model}

The key model in EKS algorithm is the state transfer model, which is composed of two parts. The first part approximates the next time step heat flux as showed below, 	
\begin{equation}\label{eq.14}
q_{k+1}=q_{k}+\omega_{q}
\end{equation}
where we consider the induced error by this approximation as a part of the noise $\omega$ applied to the state transfer model.
The second part forwardly calculates the next-time-step temperature field based on current temperature field and heat flux, which is achieved by CFD simulations in traditional methods\cite{24,25,26}, where the input of state transfer model is the current state vector $x_{k}=\left[\begin{array}{llll}
T_{k}^{1} & T_{k}^{2} \ldots T_{k}^{i} \ldots T_{k}^{N} &  q_{k}
\end{array}\right]^{T}$, and the output is the next-time-step whole temperature field $\left[\begin{array}{lll}T_{k+1}^{1}\ldots T_{k+1}^{i}\ldots T_{k+1}^{N}\end{array}\right]^{T}$

Despite the proven feasibility of this chosen state vector, the computational cost is still prohibitively expensive for online estimations. Large amounts of redundant sensitivity analysis are generated from this high-dimensional state vector containing the entire temperature field of all mesh nodes.

To address this problem, a novel state transfer model of reduced dimension is designed here with the aid of ANN algorithm. The reduced form of state vector is as follows, which serves as the input of our ANN algorithm.

\begin{equation}\label{eq.15}
x_{k}=\left[\begin{array}{llllll}
T_{k}^{n_{s}} & T_{k}^{n_{s}+a} & T_{k}^{n_{s}-a} & T_{k}^{n_{s}+b} & T_{k}^{n_{s}+b} & q_{k}
\end{array}\right]^{T}
\end{equation}
where the local temperature value of sensor location $T_{k}^{n_{s}}$ is listed on the state vector along with the temperatures of four other points close to the sensor.Correspondingly, the output of our ANN surrogate model can be organized as $\left[\begin{array}{lllll}T_{k+1}^{n_{s}} & T_{k+1}^{n_{s}+a} & T_{k+1}^{n_{s}-a} & T_{k+1}^{n_{s}+b} & T_{k+1}^{n_{s}+b}\end{array}\right]^{T}$, representing the next time step local temperature field.

\subsubsection{The generation of dataset and the training procedure of ANNs}
To train the ANN models, large amounts of numerical simulations are conducted to generate the training dataset.

The transient temperature field are calculated by solving the covering equations described in section
$2.1$ with finite volume method, which applies upwind differential scheme in $x$ direction, central differential scheme in $y$ direction, and the implicit scheme is adopted for the time marching [12]. The total mesh size is $25 \times 50$ with $\Delta x=0.04 \mathrm{~m}$ and $\Delta \mathrm{y}=0.002 \mathrm{~m}$.

\begin{figure}[h]
\centering
\includegraphics[scale=0.92]{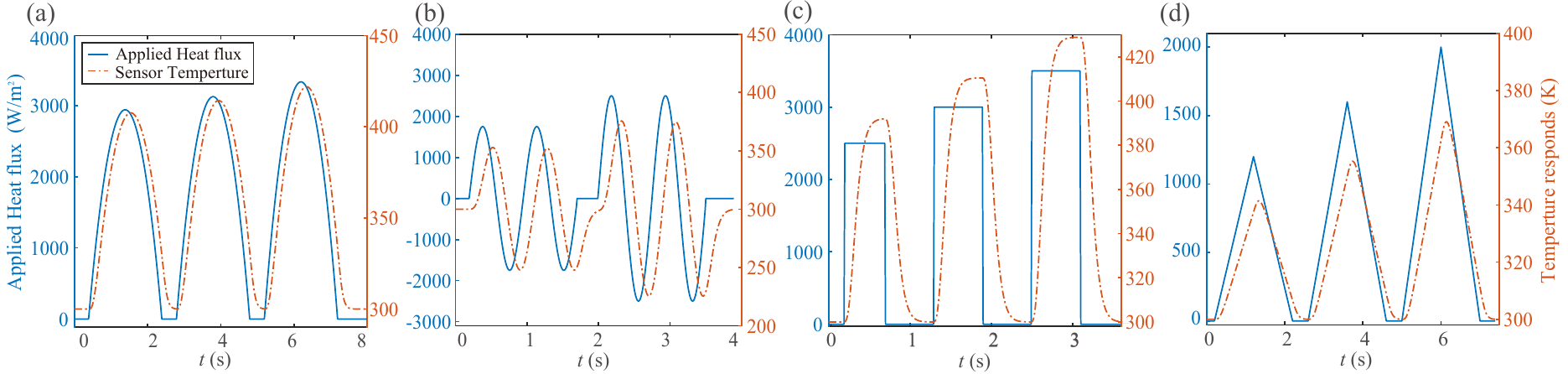}
\caption{ An extracted schematic of generating dataset by applying various form of heat fluxes and obtaining temperature distribution over time (a)   Parabolic heat flux (b) Sinusoidal heat flux (c) Step heat flux (d) Triangular heat flux} \label{fig4}
\end{figure}

To cover a variety of changing thermal boundary conditions on the heated wall, as shown in Fig. \ref{fig4}, a series of heat flux evolutions were adopted such as step functions, polynomial functions, sinusoidal waves and triangular waves in different amplitudes and frequencies, which will generate rich state transfer information for the training of ANN models. Moreover, the training dataset could be furthered expanded by applying the Eqs. \ref{eq.9} during the above simulation, so as to simulate the sensitivity analysis by elevating and lowering one of the state variable’s value and then calculating the next-time results.	

Therefore, the final training heat flux, with a total times steps of 6794 (time duration of 67.94 s on the testing computer), is applied to the benchmark problem including the four above forms of heat fluxes showed in Fig. \ref{fig4} with different amplitudes and frequencies. The CFD predicted temperature field is then reorganized into the standard form of inputs and outputs of ANN for training purpose.

The final dataset includes state transfer data needed in Eqs. \ref{eq.10} as well as the sensitivity analysis data needed in Eqs. \ref{eq.8} and \ref{eq.9}  . However, the data volume of the former is only 1/12 of the latter. The unbalanced dataset may cause the ANN model fitting more closely to the pattern of sensitivity analysis but works poorly under state transfer scenario. To resolve this unbalanced dataset problem, we choose to divide the dataset and train two ANN models used for state transfer prediction and sensitivity analysis respectively. Therefore, 6794 sets of data are used for training the state transfer ANN model and the remaining 81529 sets of data are used for training the sensitivity ANN model.

The fully connected multi-layer perception (MLP) neural network is chosen in this work. Both ANN models compose of three layers, with 10 neurons in the hidden layer. The inputs and outputs are standardized in order to enhance training performance. The hyperbolic tangent (tanh) function is chosen as the active function and the Levenberg-Marquardt backpropagation algorithm is used to train the neural network. The training procedure of ANN model for state transfer prediction ends after 1422 iterations when the performance of the network stops improving in the validation dataset for 6 consecutive epochs. To validate the training method, datasets are randomly divided into three parts, 70\% of which are used for training, 15\% for validation and 15\% for testing respectively. The pre-trained neural network eventually achieves a mean square error of 7.14×$10^{-9}$  in the testing dataset and the regression R value reaches 0.99999,while the training procedure of ANN model for sensitivity analysis ends after 552 iterations and achieves a mean square error of 8.02×$10^{-9}$ in the testing dataset and the regression R value reaches 0.99997, indicating that the training is successful and the obtained neural network can meet the accuracy criterion for surrogating the state transfer models.

\section{Results and discussions}

\subsection{Verification on the feasibility of the proposed algorithm }

\begin{figure}[h]
\centering
\includegraphics[scale=1]{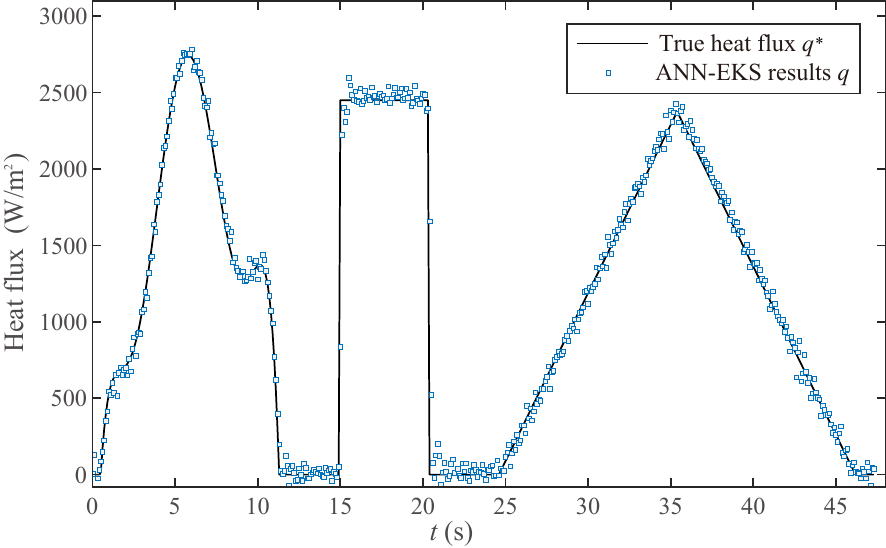}
\caption{ Comparison between the real heat flux and the estimation results under noise level m =5 K (Plot interval: every 11 time steps)} \label{fig5}
\end{figure}

In order to test the feasibility of the proposed algorithm, a transient CFD simulation is performed, in which the heat flux on the upper wall is varying in the manner showed in Fig. \ref{fig5} (solid line). The CFD predicted temperature variation at the point $(\mathrm{x}, \mathrm{y})$, is then extracted to simulate the temperature measurements with a sensor.

Given that the sensor-measured temperature is inevitably contaminated with noises, a zero-mean Gaussian noise $(\sigma \sim N(0,1))$ with a noise level $m$ is imposed on the simulated temperature evolution at the point $(\mathrm{x}, \mathrm{y})$ to simulate the noisy measurements $y_{k}$,
\begin{equation}\label{eq.18}
y_{k}=T_{k}+m \sigma
\end{equation}
where $T_{k}$ is the simulated real temperature value at sensor location of time step $k$.

In this test, the noise level is set to be m = 5. Feeding the artificial measurements $y_k$  into the ANN-EKS algorithm as the input, one can get the output $q_k$. Fig. 5 compares $q_k$ with the wall thermal boundary condition used in the CFD simulation ${q}_{k}^{*}$, which could be treated as the “true” value. The results show that the predicted heat flux evolution agrees well with the “true” value, indicating the ANN-EKS based IHTPs solver works well to predict the historical thermal boundary condition that varied with complex wave functions, such as step function, a triangular wave and an arbitrary smooth curve.

To evaluate the algorithm’s performance quantitatively, the average error ($AE$) is defined to describe the accuracy of the estimation results,
\begin{equation}\label{eq.17}
A E=\sqrt{\frac{1}{n} \sum_{k=1}^{n}\left(\bar{q}_{k}^{*}-\bar{q}_{k}\right)^{2}}
\end{equation}
where the $\bar{q}_{k}^{*}$ and $\bar{q}_{k}$ are the dimensionless form of applied heat flux and estimation results and $\mathrm{n}$ is the number of total time steps. Also, the average computing time per time step is defined to evaluate the computational efficiency of our algorithm. All the simulations involved in this work are performed on a personal computer with 2.50 GHz Intel (R) Core (TM) i7-11700F processor and 32 GB of RAM.

According to the calculation, the average error $AE$ in the baseline test is 0.0342 and average time cost is 3.649 ms per step on the testing computer. The above results show the feasibility and accuracy of our proposed algorithm.

\subsection{Effects of training dataset on prediction performance}
The offline training procedure of the ANN model is a crucial part of the proposed algorithm, so it is important to examine the effects of training dataset on the inversion performance.

This investigation is conducted in the form of ablation study. The dataset used in section 3.3 is viewed as the baseline dataset,denoted as Dataset \#0, which are generated by the training heat flux with all the four forms of varying heat flux and a total length of 67.94 s. Different types of heat flux evolutions, including step functions, triangular waves, parabolic waves and sinusoidal waves, are distributed equally in the baseline  training heat flux. Then, we gradually reduce the diversity of the training heat flux forms by directly removing certain types of the training heat flux from the baseline Dataset \#0 and then train new ANN models and test the corresponding estimation performances. The testing heat flux composes of an arbitrary smooth curve, a step function and a triangular wave, all with different amplitude and frequency from the training heat flux, to better examine the algorithm’s ability to deal with complex changing conditions in the unknown heat flux such as smoothly changing, abruptly shifting and constantly growing or plummeting. The results are summarized in Table 2.

As shown in Table 2, there are some revealing patterns that are useful for directing our training procedures. Firstly, in general, the more abundant the training dataset is, the more likely it will perform better estimation results. For instance, the baseline dataset ranks first among others in terms of estimation accuracy (evaluated with $AE$) while the training datasets missing 3/4 of heat flux types (Dataset \#11-14) perform the worst, where others generally rank in between.

Secondly, it can be induced that certain forms of heat flux missing in the training dataset may links to an increase of estimation error in the corresponding testing heat flux. For example, in Dataset \#3 and Dataset \#6, the datasets missing step functions results in a particular raise in the average error of 0.1145 and 0.2386 under the testing step heat flux. And like in Dataset \#1,2,4, there are also some corresponding increases of $AE$ in the testing smoothed curve and triangular wave when training dataset missing some of the gradually changing heat flux like sinusoidal waves and parabolic functions.

Thirdly, the results also show that our proposed algorithm has a certain level of generalizing ability. Like in Dataset \#11, \#12, the dataset containing only step functions or triangular waves can also lead to relatively accurate estimation results in other types of testing heat flux.

\begin{table}[]
 \begin{threeparttable}
 \caption{\label{tab2}Different training dataset and the corresponding $AE$ under different testing heat flux }
 \centering
\begin{tabular}{llcccc}
\hline\multicolumn{2}{l}{ Training dataset } & \multicolumn{4}{c}{ Corresponding $AE$ in different testing heat flux } \\
 \cline{3-6} No. & Exclusion\tnote{1} & Arbitrary curves & Step function & Triangle waves & Overall \\
\hline 0 & None & $0.0164$ & $0.0654$ & $0.0177$ & $0.0320$ \\
\hline 1 & Sin & $0.0188$ & $0.0715$ & $0.0198$ & $0.0336$ \\
 2 & Para & $0.0265$ & $0.0737$ & $0.0267$ & $0.0385$ \\
 3 & Step & $0.0372$ & $0.1145$ & $0.0388$ & $0.0586$ \\
 4 & Tri & $0.184$ & $0.0673$ & $0.197$ & $0.0337$ \\
\hline 5 & Sin, Para & $0.0172$ & $0.0716$ & $0.0152$ & $0.0350$ \\
 6 & Step, Para & $0.0218$ & $0.2386$ & $0.0653$ & $0.1554$ \\
 7 & Tri, Para & $0.0164$ & $0.0684$ & $0.0142$ & $0.0331$ \\
 8 & Sin, Tri & $0.0195$ & $0.0681$ & $0.0190$ & $0.0333$ \\
 9 & Sin, Step & $0.0163$ & $0.0697$ & $0.0158$ & $0.0341$ \\
 10 & Step, Tri & $0.0166$ & $0.0652$ & $0.0172$ & $0.0325$ \\
\hline 11 & Sin, Para, Tri\tnote{2} & $0.0300$ & $0.0752$ & $0.0348$ & $0.0430$ \\
\hline 12 & Sin, Para, Step & $>1$ \tnote{3}& $0.0915$ & $0.0181$ & $>1$ \\
\hline 13 & Sin, Tri, Step & $>1$ & $>1$ & $>1$ & $>1$ \\
\hline 14 & Para, Tri, Step & $>1$ & $>1$ & $>1$ & $>1$ \\
\hline

\end{tabular}
\begin{tablenotes}
     \item[1] {The column “Exclusion”  means the certain types of heat flux are removed from the baseline dataset.}
     \item[2] {“Sin”, “Para” ,“Tri” stands for sinusoidal waves, parabolic waves,  triangular waves respectively and “Step” stands for step functions.}
    \item[3] {“$>1$”means the algorithm cannot function normally and produces highly invalid results.}
   \end{tablenotes}
 \end{threeparttable}
\end{table}

To summarize, the diversity of training dataset types can lead to better estimation performance, among which the step function is relatively more important than other forms of training heat flux and are essential to enhance the algorithm’s ability to invert unknown heat flux with abrupt changing conditions. Other forms of training heat flux like triangular waves, sinusoidal waves and parabolic functions are also of additional help to the proposed algorithm to handle complex changing boundary conditions. So when constructing the training dataset, it is advisable to properly include more step functions and also increase the diversity of heat flux types to ensure better inversion performance.

\subsection{Comparison study with other algorithms}
In order to highlight the advantage of the proposed algorithm, it is compared with the inverse ANN algorithm \cite{16, 17} and the extended Kalman smoothing algorithm (EKS)\cite{24,25}.

The inverse ANN algorithm\cite{16,17} mainly utilizes the artificial neural network model to directly map the sensor-measured temperature to the unknown boundary conditions. The inputs of the ANN model are the past and future temperature measurement at the sensor locations $\left[T_{k+n_{p}} \ldots T_{k-1}\quad T_{k}\quad T_{k+1} \ldots T_{k+n f}\right]^{T}$, and the outputs are the boundary heat flux $\left[q_{k}\right]$ in the current time step. The inverse ANN algorithm possesses high computational efficiency due to its direct prediction of unknown BCs from sensor measurements, but it may tend to over-fitting and perform poorly when the measurement noises are relatively high.

The extended Kalman smoothing algorithm\cite{24,25} employs the full form of the state vector, the transfer of which is calculated based on the traditional CFD methods. Though this algorithm copes well under noisy measurement, the redundant sensitivity analysis, caused by the high-dimensionality of the full-form state vector, lead to tremendously high computational cost and limits its applications toward online estimations.

\begin{figure}[h]
\centering
\includegraphics[scale=1]{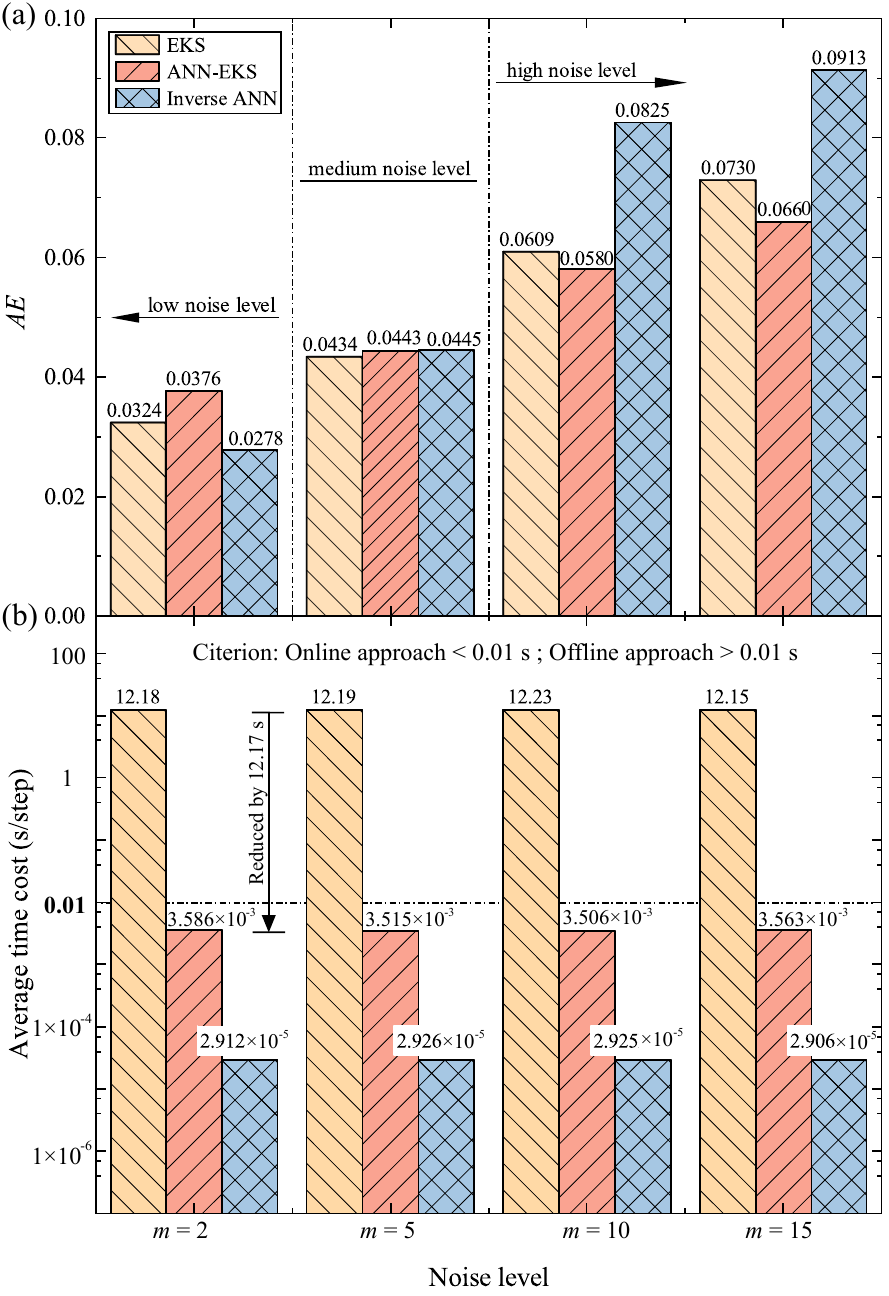}
\caption{The (a) $AE$ and (b) average time cost per step of three different algorithms under different noise level $m = 2,5,10$ and $15$ K } \label{fig6}
\end{figure}

Fig.\ref{fig6} compares the prediction precision (evaluated with $AE$) and average time cost per step of the three algorithms with a series of sensor measurements noise level $m = 2,5,10$ and $15$ K. Since the primary goal of our algorithm is online estimation, we set a criterion to evaluate the algorithm’s online ability, which is the average time cost per step should not exceed the computational time interval (0.01s in this case).

It can be seen that the proposed algorithm has a lower accuracy than the traditional inverse ANN algorithm under relatively lower noise level. However, it outperforms the inverse ANN algorithm when the noise level is higher than 10, indicating that the proposed algorithm works robustly under noisy input data. Moreover, the $AE$ of the proposed algorithm is close to that of the CFD-based EKS algorithm, which proves that the simplification of our work in the state vector is reasonable.

\begin{figure}[h]
\centering
\includegraphics[scale=1]{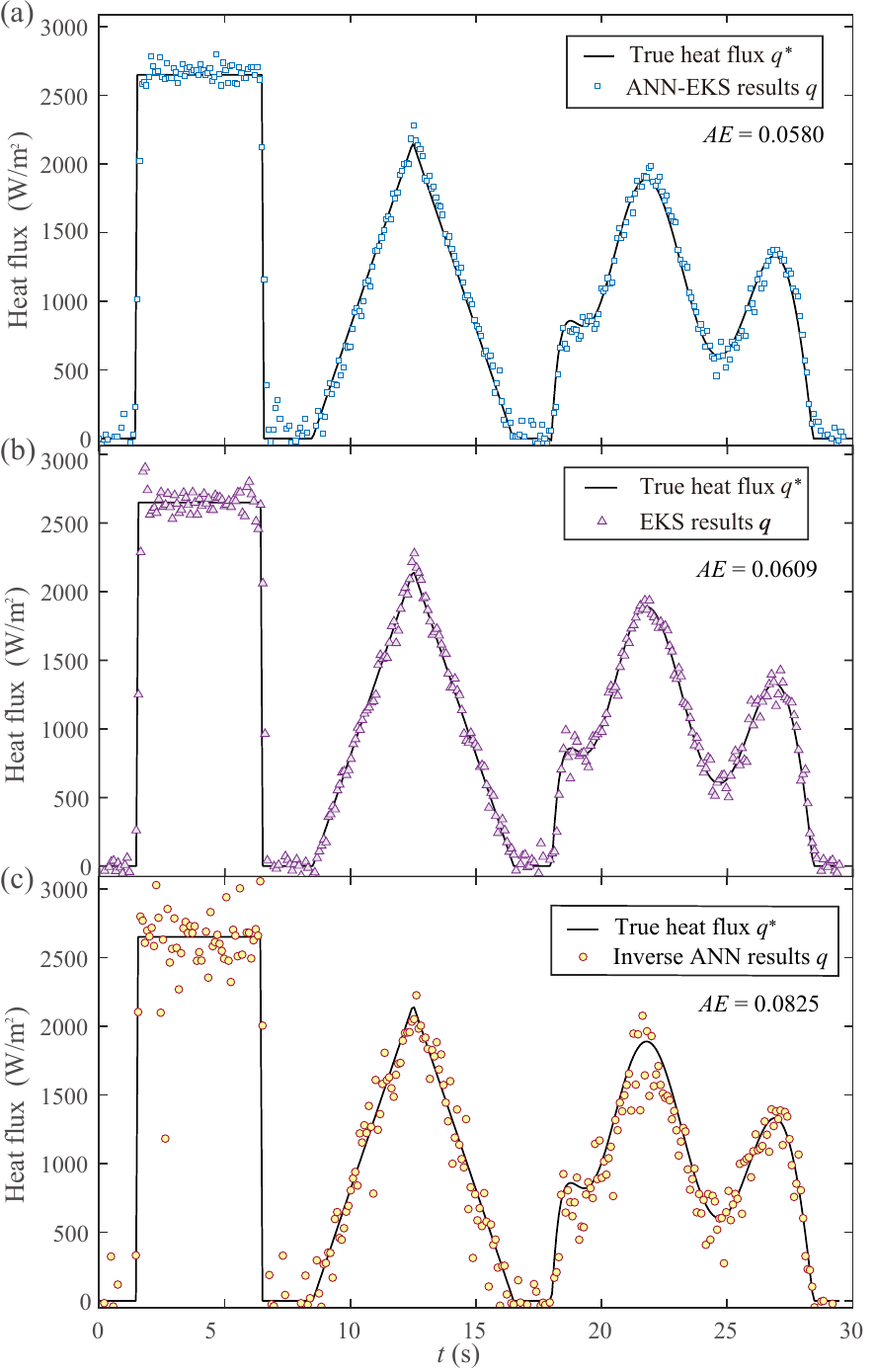}
\caption{The estimation results of the three algorithms under noise level $m = 10$ K (Plot interval: every 9 time steps) } \label{fig7}
\end{figure}

Notably, Fig. \ref{fig7} shows the inversion results of the above three algorithm (scatter) against the testing heat flux (solid line) under a relatively high noise level of $m = 10$ K. The results of our ANN-EKS algorithm and traditional EKS algorithm match the real heat flux well with an $AE$ of 0.0580 and 0.0609 respectively while the inverse ANN algorithm seemingly oscillates greatly under this high noise level and performs the worst with an $AE$ of 0.0825. Also, we can notice that the proposed ANN-EKS algorithm possesses better ability to suppress overshooting and oscillation of the prediction compared with the other two methods, which further proves the robustness of our algorithm in noisy environments.

More importantly, it can be clearly seen from Fig. \ref{fig6} that our proposed algorithm has great advantages in terms of computational efficiency. The ANN-EKS algorithm can drastically reduce the computational time of the conventional EKS approach from 12.23s per time step to 3.506ms. Judged by the criterion for online estimation, 3.506ms is significantly shorter than the time step interval of 0.01s, which means our algorithm is fully capable of performing online estimation task while the conventional EKS approach cannot.
Based on the comparison results, it is safe to say that our proposed algorithm is indeed a robust and rapid approach and capable of achieving online estimation task for solving IHTPs.

Based on the comparison results, it is safe to say that our proposed algorithm is indeed a robust and rapid approach and capable of achieving online estimation task for solving IHTPs.

\clearpage 
\clearpage

\subsection{Effects of future time steps and sensor locations on inversion performance}
There are several hyper-parameters that can strongly influence the performance of the proposed algorithm, among which the sensor locations and the future time step $n_{f}$ are investigated in this section, so as to choose these parameters wisely and guide our algorithm to reach its optimal performance.
\subsubsection{The sensor location}
The sensor location plays an significant role in the estimation accuracy. To examine its effect in detail, different possible locations of the sensor are traversed to obtain the corresponding estimation results.

\begin{figure}[h]
\centering
\includegraphics[scale=1]{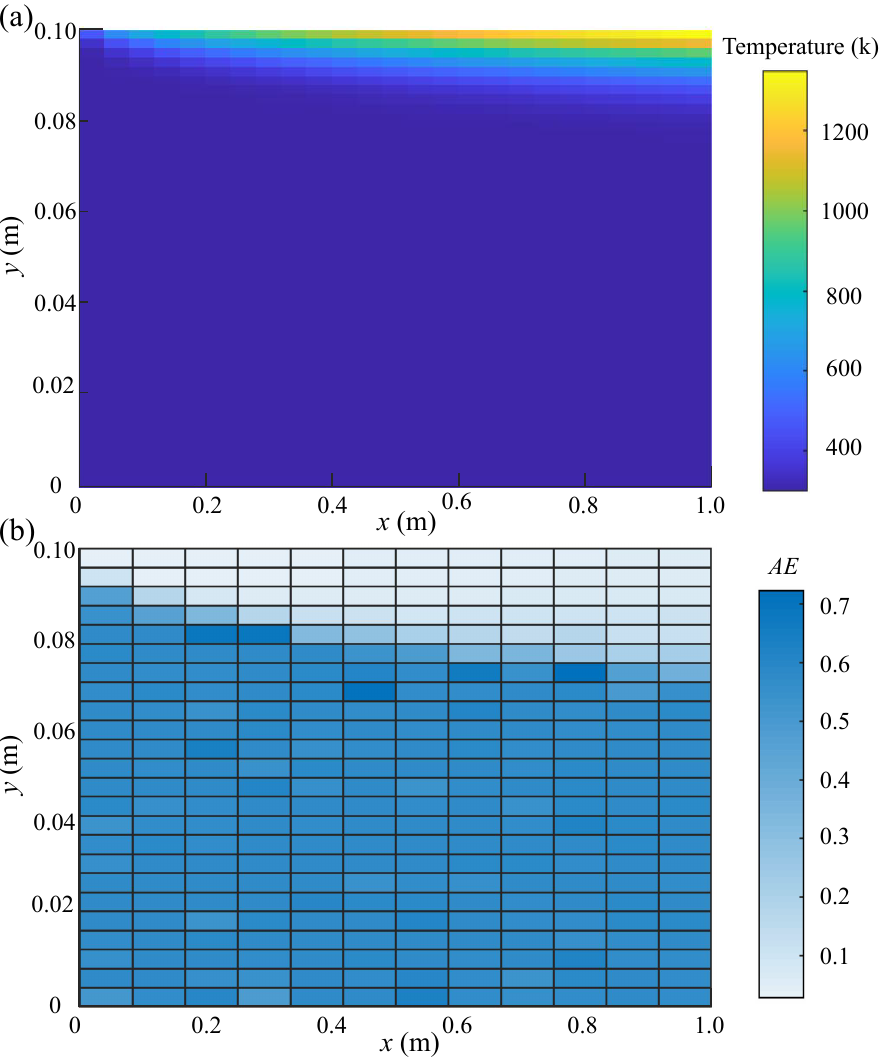}
\caption{Contour for temperature distribution in the benchmark problem and the estimation error heatmap under different sensor locations.(a) The contour map is used for demonstration, drawn when the benchmark problem have been applied with a constant heat flux of $2500 \mathrm{~W} / \mathrm{m^2}$ for $2 \mathrm{~s}$. (b)The corresponding estimation AE heat map when sensor is placed in different locations throughout the whole computational domain. The lighter one position is, the better estimation results can be obtained via our algorithm under that exact sensor locations. } \label{fig8}
\end{figure}

Fig. 8a contours the transient temperature field of the benchmark problem when heated with a constant heat flux of 2500 W/s for 2 s, which shows the general temperature distribution pattern under upper boundary heat flux. Fig. 8b shows the corresponding estimation accuracies (evaluated with $AE$) of our ANN-EKS algorithm when sensor is placed in different locations.

The results in Fig.8 demonstrates a clear link between real physic (temperature contour) and inversion performance ($AE$ heatmap). Firstly, sensor locations that are closer to the upper boundary , which the unknown heat flux are applied to, tends to perform better. Secondly, following the direction of air flow, the downstream locations also achieve relatively higher estimation accuracy compared with upstream one, forming a clear average error “boundary layer” that much resembles the heat propagation pattern from the upper boundary into the fluid field showed in Fig. 8a.

This exhibited pattern shows that we can view the inversion procedure as restoring the unknown original energy input (heat flux in this case) from sensor measurement after it has been distorted and weakened in the propagation pathway. The closer or more downstream the sensor is to the upper boundary, the more clear and less decayed information will it receive to give a better inversion results. And because of the limited ability for the heat flux to penetrate into the fluid field, the sensor been placed in further locations (dark blue areas in Fig. 8b) cannot receive enough information and thus produces highly invalid estimation results.

Therefore, in real applications, it is advisable to place the sensor in the closest possible location along the direct pathway of heat propagation in order to obtain better estimation performances.

\subsubsection{The future time step $n_{f}$}

From the former investigation on sensor locations, we can know that the spatial distances between unknown BCs and sensors have significant effect on the final inversion accuracy. Naturally, it can be assumed that the temporal effect of sensor measurement cannot be ignored as well, which, in this case, can be represented by the future time step $n_{f}$ in our RTS smoothing algorithm.

In order to examine the influence of $n_{f}$, numerical experiments with different values of $n_{f}$ from 0 to 30 are conducted, in which case the sensors are placed in several plausible locations. The resulting performance of our proposed algorithm, in terms of the average error and the computational cost per time step is depicted in Fig.9.

\begin{figure}[h]
\centering
\includegraphics[scale=1]{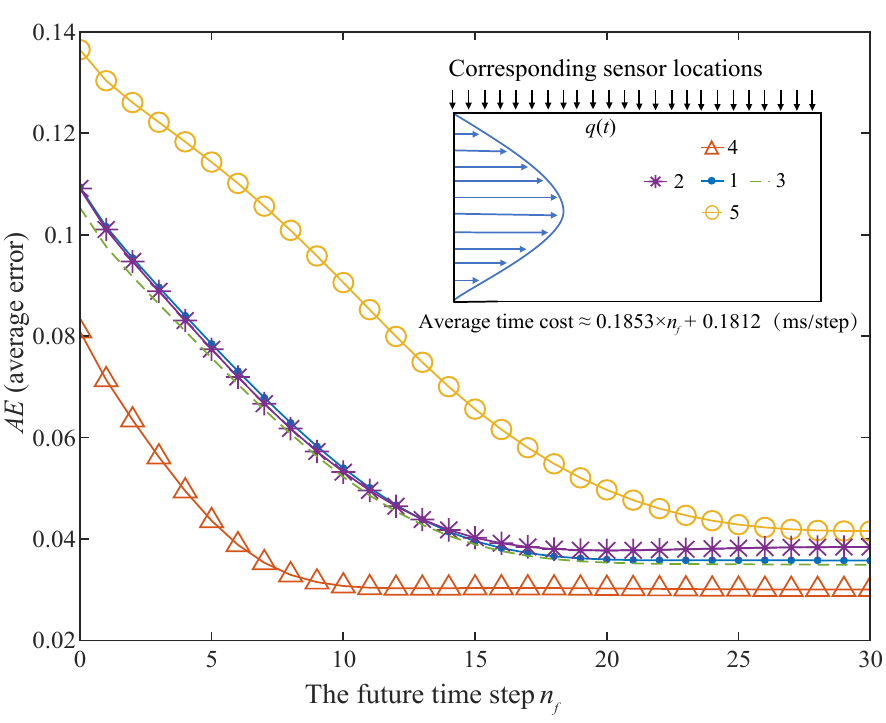}
\caption{The performance of the proposed algorithm under different $n_{f}$ and sensor locations. Sensor \#1-5 are respectively located at $(0.820,0.091),(0.940,0.091),(0.700,0.091),(0.820,0.091)$, $(0.820,0.087)$. The noise level $\mathrm{m} = 5 \mathrm{~K}$ and the training dataset is the baseline dataset. Other parameters remains the same with previous sections. Due to the consistency of the computational cost under different sensor locations, a linear expression of the average time cost with respect to $n_{f}$ is obtained.} \label{fig9}
\end{figure}

In general, all the five sensors placed in different locations show consistent patterns. The average computational cost of for all the five sensors grows linearly with respect to the future time step $n_{f}$ while their corresponding estimation $AE$s decline at first and gradually approach to a stable value when the $n_{f}$ reaches a certain threshold.

\begin{figure}[h]
\centering
\includegraphics[scale=1]{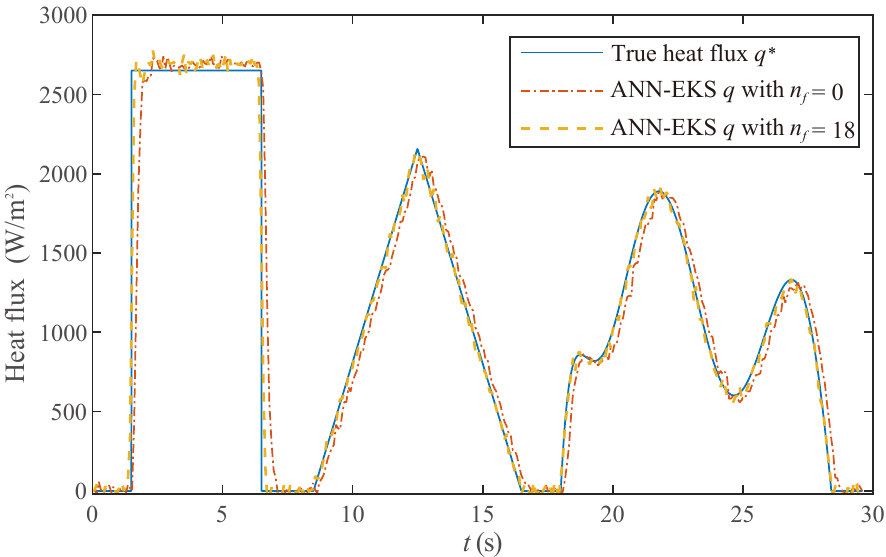}
\caption{ The estimation results of our proposed algorithm with $n_{f}=0$ and $n_{f}=18$. The noise level $\mathrm{m}=5 \mathrm{~K}$ and the sensor is placed at $(0.820,0.091)$ and the training dataset is the baseline dataset. Other parameter remains the same with previous sections.} \label{fig10}
\end{figure}

The temporal effect of sensor measurement exhibits clearly on the varying pattern of AE with respect to different future time steps $n_f$. As showed in Fig. \ref{fig9}, the lowest estimation accuracy for all the five sensors are achieved under $n_f=0$. When $n_f=0$, our ANN-based extended Kalman smoothing (EKS) algorithm degrades into the ANN-based extended Kalman filtering (EKF) algorithm, which does not utilize the future measurement to address the sensing delay issue and thus may obtain a lagged solution. To illustrate this effect, Fig .\ref{fig10} is depicted to compare the estimation results when $n_f=0$ (dashdot line) and $n_f=18$ (hidden line), where the estimation results under $nf=0$ experience obvious lagging problem compared with the real heat flux or results under relatively large $n_f$.

Then, with the increase of future time step $n_f$, higher estimation accuracies are obtained as showed in Fig. \ref{fig9}, which may because the sensing delay issues are gradually lessened by utilizing more future time measurement (larger $n_f$). However, beyond a certain threshold of $n_f$, the growth in the estimation accuracy seems stagnant, indicating that the sensor delay issue has been well addressed and thus no further improvement of the estimation accuracy can be achieved by introducing more future measurement. The results also shows that different sensor locations correspond to different threshold, with the closest sensor 4 needing about 10 future time steps to reach its optimal estimation accuracy while the the farthest sensor 5 needing about 28 future time steps. The reason behind this phenomenon is that the unknown boundary heat flux may firstly penetrate to closer sensors, which may experience lighter sensing delay problem and thus require less amount of future measurement to achieve relatively higher estimation accuracy.

On the other hand, the spatial effect of sensor locations on the inversion performance can be noticed as well. From Fig. \ref{fig9}, closer locations toward the upper boundary in$y$ direction tends to produce higher estimation accuracy and downstream positions of sensors also produce marginally higher estimation accuracy compared with upstream positions, which is consistent with the pattern we discovered in the previous section. Also, we can see that the spacial effect along the flow is not as obvious as that of $y$ direction, which indicates that the heat conduction along $y$ direction is more intense than the convective heat transfer along the flow in the benchmark problem.

Therefore, it can be induced that a too small future time step leads to lagness problem with high estimation error while a too large time step requires more computing time but without much improvement in the estimation accuracy. Thus the future time step $n_{f}$ should be chosen via the appropriate trade-off between the estimation accuracy and the computational efficiency while also taking the spatial effect of sensor locations into consideration.

\section{Conclusion}
A rapid yet robust inversion algorithm, ANN-based extended Kalman smoothing algorithm (ANN-EKS), is developed to realize the online estimation of time-varying thermal boundary conditions in  a two-dimensional tube convective heat transfer problem. The major findings are summarized as follows:

(1) The proposed algorithm is a computational-light online approach for the estimation of the unknown boundary conditions. Compared with conventional CFD-based EKS algorithm, the computational costs of the proposed are reduced drastically from 12.23 s per time step to 3.506 ms, which makes our algorithm fully capable of performing online estimation task.

(2) The proposed algorithm is relatively robust to handle measurement data with high noise-signal ratio. An average error of 0.0580 for estimating unknown boundary heat flux can be achieved via our algorithm, whose accuracy is basically equivalent to the conventional EKS algorithm with an average error of 0.0609 and improves significantly compared to the inverse ANN algorithm with an average error of 0.0825.

(3)  ANN model is employed as a surrogate model to predict the state transfer process of given thermal system. It is trained with unsteady CFD simulation data with a variety of time-dependent thermal boundary conditions of tube wall, such as step function heat flux, triangular wave heat flux, sinusoidal wave heat flux, and parabolic heat flux, among which the step function is more crucial than other forms of data, because in ablation study the most cases that missing step functions experience relatively larger growth in the estimation error from 0.0320 to 0.0586 of Dataset \#3 and 0.1554 of Dataset \#6.

(4) The sensor location and the future time steps $n_{f}$ are two coupling influential factors to the prediction of the wall thermal boundary condition. Spatially, a better prediction is realized while the sensor is moved towards downstream or close to the wall, indicating that the design of sensor location should consider the propagation pathway of the thermal perturbation from the wall. Temporally, relatively larger future time steps $n_{f}$ can ensure the algorithm to better cope with the sensing delay issues, indicating that the choice of future time steps $n_{f}$ should also consider the temporal lagging pattern from the applied boundary heat flux to the sensors.

\section*{Acknowledgements}
Project 51806005 supported by National Natural Science Foundation of China
\nocite{*}

\section*{Data Availability Statements}
The data that support the findings of this study are available from the
corresponding author upon reasonable request.
\nocite{*}


\end{document}